\documentclass[reqno]{amsart}
\usepackage{oldlfont}
\usepackage{citesort}
\usepackage{amssymb}
\newtheorem*{ack}{Acknowledgments}
\begin{document}
\title{On reaction thresholds in doubly special relativity}
\author{Dan Heyman$^\dag$, Franz Hinteleitner$^{\dag,\S}$, and Seth Major$^{\dag}$}
\date{December 2003}
\address{$^\dag$Department of Physics\\
Hamilton College\\
Clinton NY 13323 USA\\ 
$^\S$Department of Theoretical Physics and Astrophysics Masaryk University\\
Kotl\'{a}\v{r}sk\'{a} 2, 611 37 Brno, Czech Republic}
\email{smajor@hamilton.edu,franz@physics.muni.cz}

\begin{abstract}
Two theories of special relativity with an additional invariant scale, ``doubly special relativity" (DSR), are tested with calculations of particle process kinematics.  Using the Judes-Visser modified conservation laws,  thresholds are studied in both theories. In contrast to some linear approximations, which allow for particle processes forbidden in special relativity,  both the Amelino-Camelia and Magueijo-Smolin frameworks allow no additional processes.  To first order, the Amelino-Camelia framework thresholds are lowered and the Magueijo-Smolin framework thresholds may be raised or lowered.
\end{abstract}

\maketitle

\section{Introduction}

Special relativity with an observer independent scale has been proposed as a modification to local Lorentz invariance \cite{AC,ISR1trans,AC02rev,ACnature,ACISR3,MS1,MS2}. The existence of an additional scale at high energy was motivated by a variety of studies including $\kappa$-deformed Poincar\'e algebras \cite{ISR1trans,ACM,LN,KGN,KGN2,KR}, heuristic semi-classical states in quantum gravity \cite{AM-TU},  and string theory \cite{stringy}. The new scale may be an energy, momentum, or perhaps even a length. Despite our intuition from special relativity, the new relativity theories seem to demonstrate that it is {\em not} necessary to use a preferred reference frame when there is a distinguished scale \cite{AC}. Dubbed ``doubly special relativity" (DSR)  the theories maintain the relativity principle even with the inclusion of an invariant energy or momentum \cite{AC}. For the purposes of this paper, the distinguishing features of the new theories are the relativity principle and an invariant scale.  To emphasize this we refer to them as ``invariant scale relativity" (ISR).  In ISR theories the speed of light may not be an observer invariant.\footnote{For instance, the modified dispersion relation $E^2=p^2 + p^2 E /E_p$ yields a velocity of \cite{AC} $$v_\gamma(p) := \frac{dE}{dp} \approx 1+\frac{p}{E_p}$$ which depends on the reference frame for $p \neq E_p$.} We study two example theories, the ISR of Amelino-Camelia and collaborators
\cite{AC,ISR1trans,AC02rev,ACISR3} and the ISR of Magueijo and Smolin
\cite{MS1,MS2}.  Both proposals exploit a freedom to define non-linear transformations on momentum space, retaining the group properties of Lorentz transformations, and include an invariant scale. 
 
Defined in momentum space the new ISR transformations raise many questions. For instance, is the relativity principle maintained? Indeed, what is the relativity principle in this new context? What is the corresponding spacetime associated with these theories?\footnote{At the present, despite some progress \cite{KMM}, it is unclear precisely how this scale affects relativistic effects such as length contraction.}  How are composite particles described?  Using particle process kinematics to test relativity in the ISR models, we focus on the first two questions and, to the extent possible, limit ourselves to the single particle sector.

Studies of process kinematics, together with current astrophysical observations, have been surprisingly successful in constraining specific proposals for modifications of special relativity requiring a preferred frame \cite{KM,JLM,JLMsynch,JLMS}.  Thus far these studies have focused on modifications of dispersion relations with a term linear in the Planck scale.    Further constraints may be imposed by ensuring consistency at lower energies via an effective field theory, as was done for dimension five operators by Myers and Pospelov  \cite{MP}.  Lehnert found constraints on dispersion relations arising from the additional considerations of coordinate invariance and non-dynamical tensor backgrounds  which break Lorentz symmetry \cite{RL}.   

Kinematics is particularly well suited to non-linear realizations of the Lorentz group since both the spacetime picture and the effective dynamical framework of ISRs is  not complete.  To perform the
analysis we need conservation laws.  Judes
and Visser derive modified conservation laws in Ref. \cite{JV} based on the observation that, since the
physical energy-momenta in ISRs is non-linearly related to the
formal energy-momenta, the ISR conservation laws may be found by
appropriately applying the non-linear transformations to the usual
additive conservation laws.  

Given the success constraining modified dispersion relations in
Refs. \cite{KM,JLM,JLMsynch,JLMS}, we might expect that process kinematics could
again be used to constrain the new invariant scale in ISRs. In
fact, although this is the first general study, several
such process, including photo-production of pions occurring in high-energy proton - cosmic microwave background photon collisions (the Greisen, Zatsepin
and Kuzmin (GZK) cutoff \cite{GZK}), have been explored \cite{MS2,ACISR3}. These
calculations have been carried out in the leading order formalism.
Here, making use of the Judes-Visser conservation laws, we present
new, exact and first order calculations for the Magueijo-Smolin and Amelino-Camelia ISRs.
Particle process kinematics does not limit parameters in the same manner as Refs. \cite{KM,JLM,JLMsynch,JLMS}. Instead, process kinematics shows how thresholds are modified and provides a perspective from which the notion of relativity may be sharpened.
Indeed particle kinematics brings matters of principle to the fore
in ISRs rather than numerical limits on parameters.

We present our results for Magueijo-Smolin ISR before
turning to Amelino-Camelia ISR in Section \ref{ISR1}. We show that to first order Amelino-Camelia ISR lowers existing thresholds, whereas Magueijo-Smolin ISR may either lower or raise them. 
 They allow no additional processes. We explore the issue of the uniqueness of particle 
process thresholds in Section \ref{uniqueness} and close with a brief discussion of the relativity principle in light of these results.

Throughout the article when we refer to the ``Planck scale" we simply
mean the invariant scale of the theory expected to correspond to $E_p = 
1.3 \times 10^{19}$ GeV. The low-energy speed of light is set to unity.  We generally calculate in $1+1$ for simplicity.  However, in section \ref{uniqueness} where the results depend on dimension, we work in $3+1$. 

\section{Magueijo-Smolin's relativity with an invariant energy}
\label{ISR2}

Fock, in {\em The theory of space-time and gravitation} \cite{fock},
derives spacetime transformations for a system in which linear motion
is covariant; if motion is rectilinear in one frame, then is
rectilinear in all inertial frames.  He showed that the
transformations from a frame $x^\mu$ to $x^{\mu \prime}$ must be of
the form \begin{equation} x^{\mu \prime}=\frac{A^\mu+A^\mu_\nu
x^\nu}{B+B_\alpha x^\alpha} \end{equation} where $A^\mu, A^\mu_\nu,
B,$ and $B_\mu$ are coordinate independent functions of velocity. 
Magueijo and Smolin found that these same transformations applied in
momentum space introduce an invariant scale at high energy.  They
showed that the fractional linear transformations may be obtained by
exponentiation of boost generators modified by a dilation $D=p_\nu \partial_p^\nu$ \cite{MS1}
\begin{equation}
\label{MSgen}
K^i=L^i +\lambda p^i D
\end{equation}
in which $L^i$ is the unmodified Lorentz generator.

The resulting Magueijo-Smolin ISR may be defined by the physical energy-momenta for a
single particle \cite{MS1,JV}
\begin{equation}
\begin{split}
\label{ISR2def}
E = \frac{\epsilon}{1+\lambda \epsilon} \\
p =\frac{\pi}{1+\lambda \epsilon}
\end{split}
\end{equation}
and the modified dispersion relation
\begin{equation}
\label{ISR2disp} \frac{E^2 - p^2}{(1-\lambda E)^2} = \mu^2 \equiv
\frac{m^2}{(1-\lambda m)^2}.
\end{equation}
The quantities $(\epsilon,\pi)$, called ``pseudo-energy-momenta,"
transform under the usual linear Lorentz transformations.

The presence of the pseudo-energy-momentum variables in the background does not necessarily mean that the ISR trivially reduces to SR.  An ``ISR physicist" would not measure -- perhaps not even calculate -- the pseudo-energy-momentum variables. We assume that the non-linearly realized variables are the physical ones. For notational convenience we use $E_p=1/\lambda$ but this in no way is
meant to suggest that there is an invariant length.  Until the spacetime picture is complete we cannot be sure how the invariant scale relates to a possible length.

For many particle processes the total physical energy is given by the
same expression although $(\epsilon,\pi)$ becomes the total
pseudo-energy-momenta $(\epsilon_{tot},\pi_{tot})$.\footnote{As is
clear from the definition, we study Magueijo-Smolin ISR ``classic" of \cite{MS1}
rather than later variants which contain more than one scale
\cite{MS2}.} Thus, equations (\ref{ISR2def}) also define modified
energy-momentum conservation laws which, unlike the
pseudo-energy-momenta, are not additive \cite{JV}.  

Before exploring process kinematics it is worth reviewing a couple
of results on the invariant scale.  As shown in Ref.  \cite{MS1}, the
theory has an invariant energy, $E_p$, such that if a particle has
this energy in one frame then it has the same energy in all frames
(despite the change in momentum).  The Magueijo-Smolin theory also has
invariant ``Planck scale null vectors" $(E_p,\pm E_p)$.  Interpreting $E_p$ as the invariant energy, we always take $\lambda>0$.  One might wonder whether the distinguished energy is included in the momenta space accessible to physical particles.  Kinematic calculations
suggest that it should not be included.

The root of the issue is the singularity in the pseudo-energy $\epsilon=\frac{E}{1-\lambda E}$ at $E=E_p$ where ``anything can happen."  By
modified energy conservation, the total energy of $N$ particles is
\begin{equation}
E_{\rm tot} = \frac{ \sum_{i=1}^{N} \frac{E_i} {1-\lambda E_i} } {1 +
\lambda  \sum_{i=1}^{N} \frac{E_i}{1-\lambda E_i}  }=\frac{1}{\lambda}
\left[1-\frac{1}{1+\lambda\sum_{i=1}^N \frac{E_i}{1-\lambda E_i}}\right].
\end{equation}
This is always smaller than $E_p=1/\lambda$, as long as all the $E_i$ are smaller
than the Planck scale energy. If one of the $E_i$ is equal to $E_p$, then also the total energy
is $E_p$, regardless of the number of particles and the values of the other, sub-Planckian energies.

Further curiosities appear for composite particles.  Kinematically, a Planck-scale particle can decay to $N$ particles (with $N$ finite) as long as one of them has Planck-scale energy.   One may similarly check that momentum is conserved.  Indeed the derivation holds for the Planck scale null vector as well.  (See Refs. \cite{MS2,AC02rev,SU} for further complications in defining composite particles.) Thus, a
Planck-scale particle is a source (or sink) for an arbitrary number of
particles with energies less than or equal to the Planck scale.  In
addition, one may show that a finite number of sub-Planckian particles
cannot interact to produce a Planck scale particle.  Because of this
closure property for $E<E_p$ particles under process kinematics and
the pathologies of including these invariants in the physical
energy-momentum space, we take Magueijo-Smolin ISR to be defined on the space of
4-momenta satisfying the modified conservation laws and $E<E_p$. (This is analogous to what is done in SR for infinite energies.)

Process kinematics is considerably simplified by the observation
that conservation of the physical energy and momentum is
equivalent to conservation of the pseudo-energy-momenta. To see
this, consider an $M$ to $N$ particle process, with incoming
pseudo-energy 
$$\epsilon_{o} = \sum_{i=1}^M \epsilon_i =
\sum_{i=1}^M E_i/(1-\lambda E_i)$$
and outgoing pseudo-energy
$\epsilon_{f} = \sum_{j=1}^N \epsilon'_j$, energy conservation
$E_o = E_f$ then requires
\begin{equation}
\frac{\epsilon_{o}}{1+\lambda \epsilon_{o}} =
\frac{\epsilon_{f}}{1+\lambda \epsilon_{f}}
\end{equation}
which immediately implies that the total pseudo-energy is conserved.  This in turn implies that the pseudo-momentum is conserved.  However note in particular that this result does not imply that the ISR results are identical to the results of SR kinematics.  
Further, the result is by no means generic to all ISRs but a simple consequence of the
fractional modification. For instance, one might try a ``time reversal"
invariant theory with the modifications of the form $(1+(\lambda
\epsilon)^2)^{-1}$. The above argument obviously fails for such an ISR.

To compare process thresholds of the Magueijo-Smolin ISR with those of SR, we take the reaction of two incoming particles with masses $m_1$ and $m_2$, resulting in $N$ outgoing particles with masses $m_i$, $i=3,\ldots,N+2$ in the center-of-mass (CM) system. Let $M := \sum_{i=3}^{N+2}m_i$ and $M^{(2)} = \sum_{i=3}^{N+2}m_i^2$. Recall that the usual SR threshold in the CM system is given by
\begin{equation}\label{ESR}
E_{\rm SR}^*=\frac{m_1^2-m_2^2+M^2}{2M}.
\end{equation}
To find the ISR threshold the physical energies and masses in (\ref{ESR}) are replaced by the corresponding pseudo-quantities,
\begin{equation}\label{E2}
\frac{E_{\rm ISR}^*}{1-\lambda E_{\rm ISR}^*}=\epsilon^*=\frac{\mu_1^2-\mu_2^2+\mu^2}{2\mu}.
\end{equation}
with $\mu:=\sum_{i=3}^{N+2}\mu_i$. From this we obtain $E_{\rm ISR}^*$ in terms of the ISR invariants $\mu_i=\frac{m_i}{1-\lambda m_i}$ and, after expansion with respect to $\lambda$, the first-order correction of the SR threshold energy,
\begin{equation}
E_{\rm ISR}^*\approx E_{\rm SR}^*\left[1-\lambda\left(E_{\rm SR}^*-\frac{4M(m_1^3-m_2^3)-2M^{(2)}(m_1^2-m_2^2)+2M^2M^{(2)}-M^4}
{2M(m_1^2-m_2^2+M^2)}\right)\right].
\end{equation}

In the case of equal in-going masses, $m_1=m_2$, this simplifies to
\begin{equation}\label{EMS}
E_{\rm ISR}^*\approx E_{\rm SR}^*+\lambda\frac{2M^{(2)}-M^2}{4}.
\end{equation}
The sign of the correction is not generally definite, it depends on the values of the outgoing masses. In the case of two outgoing particles, nevertheless, the threshold is always raised, as (\ref{EMS}) reduces to
\begin{equation}
E_{\rm ISR}^*\approx E_{\rm SR}^*+\frac{\lambda}{4}(m_3-m_4)^2.
\end{equation}
This is not a generic result for the reaction of two different incoming particles, as we will see below.

An interesting example is the interaction of an ultra-high energetic proton from cosmic radiation with the cosmic microwave background, $p\gamma\rightarrow p\pi$, in which the proton loses energy to produce a pion. We assume in the following that, however physical momenta are defined for the composite proton and pion, the result is well-approximated by the dispersion relation for an elementary particle. The SR threshold for this process leads to a cutoff in the cosmic particle spectrum, the GZK cutoff \cite{GZK}. Recently, higher energy cosmic particles have been reported. To check whether the Magueijo-Smolin ISR could account for a raising of this threshold we specialize the above method.
From equation (\ref{ESR}) the special relativistic threshold is
\begin{equation}
\label{SRGZKCMthresh}
E_{\rm SR}^*=\frac{(m_p+m_\pi)^2+m_p^2}{2(m_p+m_\pi)}.
\end{equation}
In the Magueijo-Smolin ISR the corresponding relation is
\begin{equation}
\epsilon^*=\frac{(\mu_p+\mu_\pi)^2+\mu_p^2}{2(\mu_p+\mu_\pi)},
\end{equation}
from which follows
\begin{equation}
E_{\rm ISR}^*=\frac{(\mu_p+\mu_\pi)^2+\mu_p^2}{2(\mu_p+\mu_\pi)+\lambda[(\mu_p+\mu_\pi)^2+\mu_p^2]}.\end{equation}
In first order in $\lambda$ this is
\begin{equation}
E_{\rm ISR}^*\approx E_{\rm SR}^*-\lambda\,\frac{m_\pi^2(6m_p^2-m_\pi^2)}{4(m_p+m_\pi)^2},
\end{equation}
a {\em lowering} of the SR threshold energy in the CM system. To compare this with the GZK threshold in the cosmological frame, one performs a non-linear Lorentz transformation, which boosts $E_\gamma$ to the energy of a far infrared background photon. This is done in Appendix A. However, like in ordinary Lorentz transformations, the boosted energy is a monotonic function of the original one and so Magueijo-Smolin ISR is not capable of raising the GZK threshold and explaining the apparent abundance of cosmic particles above the GZK cutoff \cite{MS2}.  

We exhibit two exact kinematic calculations for the Magueijo-Smolin ISR in Appendix A.  These are based on two processes of the basic QED vertex, vacuum \v{C}erenkov radiation (V\v{C}R) $a \rightarrow a \, \gamma$ for a charged particle $a$ and photon stability $\gamma \nrightarrow e^+e^-$ . These processes, both forbidden in SR, are of particular interest, because considerations of linear modifications of SR \cite{KM,JLM} indicate that they could be allowed in modified theories. From the exact calculations it follows that they are forbidden in the ISR as well.

It is no surprise that we obtain these results.  For, the Magueijo-Smolin theory does not admit additional kinematic solutions. The crux of the matter is the equivalence of the conservation of the
physical energy-momenta and the pseudo-energy-momenta.  Since the map between physical energy-momentum thresholds and pseudo-energy-momentum thresholds is one-to-one, the theory contains no additional solutions (see Section \ref{uniqueness}). If a process is forbidden in SR it will remain forbidden in the Magueijo-Smolin ISR. 

\section{The Amelino-Camelia relativity with an invariant momentum}
\label{ISR1}

The next ISR we consider differs from the Magueijo-Smolin theory in a number of important ways. 
First, the Amelino-Camelia ISR does not simply contain a dilation in momentum space but
represents a more drastic modification.  This can be easily seen by comparing equation (\ref{MSgen}) to the first order form of the modified boost generators for Amelino-Camelia ISR \cite{AC02rev}
\begin{equation} 
K^i = L^i + \lambda \left( \frac{1}{2} \eta^{\mu\nu}
p_\mu p_\nu x^i+ p^i p_j x^j \right).
\end{equation} 
The dilation is only on the 3-momenta and the
non-linear action extends to the spacetime transformations.  As a result of these non-linearities, it is often necessary to work with the physical energy momenta to obtain exact results for process kinematics.
Second, the Amelino-Camelia ISR has a single invariant momentum $p_o = 1/ \lambda$ but the
energy, as in SR, is unconstrained.  The theory may again be defined
by the relation to the pseudo-energy-momenta \cite{JV}
\begin{equation}
\begin{split}\label{ISR1def}
E &= \frac{1}{\lambda} \ln \left[ 1 + \lambda \epsilon \sqrt{1 +
\frac{\lambda^2(\epsilon^2-\pi^2)}{4}} +
\frac{\lambda^2(\epsilon^2 - \pi^2)}{2} \right] \\
p &= \pi e^{-\lambda E} \sqrt{1 +
\frac{\lambda^2(\epsilon^2-\pi^2)}{4}} .
\end{split}
\end{equation}
The theory has a modified dispersion relation \cite{JV}
\begin{equation}
\label{ISR1disp} \cosh(\lambda E) = \cosh(\lambda m) + \frac{1}{2}
\lambda^2 p^2 e^{\lambda E}.
\end{equation}
This dispersion relation, to leading order \cite{AC}, is identical to the modified
dispersion relations studied in \cite{KM}. However, in the ISR
context the energy-momentum conservation laws are modified as
well \cite{AC,JV}.

As may be swiftly seen from the dispersion relations of equation (\ref{ISR1disp}), although there is an invariant momentum, no positive energy particle may obtain it. We consider only those particles with momentum less than the upper limit $p_o$. In the following we analyze the theory defined by equations (\ref{ISR1def}) and (\ref{ISR1disp}), the Judes-Visser conservation laws \cite{JV}, and the restriction $p<1/\lambda$.  For ease of reference we will refer to this theory as Amelino-Camelia ISR.

The calculation of leading order corrections to threshold energies in the CM frame begins with the observation that the invariant $\mu$ of the theory differs only in second order from the physical mass,
\begin{equation}
\mu=\frac{2}{\lambda}\sinh\frac{\lambda m}{2}\approx m+\lambda^2\frac{m^3}{24}.
\end{equation}
From this it follows that the threshold pseudo-energy for a general $2\rightarrow N$ particle process, given by the right equality of equation (\ref{E2}), is
\begin{equation}
\epsilon^*=E_{\rm SR}^*+O(\lambda^2),
\end{equation}
which greatly simplifies the calculation of the first order expression of the threshold energy $E_{\rm ISR}^*$ in Amelino-Camelia ISR.  With the aid of equation (\ref{ISR1def}),
\begin{equation}
E_{\rm ISR}^*\approx E_{\rm SR}^*-\frac{\lambda\pi_1^2}{2}.
\end{equation}
Here $\pi_1$ is the pseudo-momentum of the in-going particle, whose pseudo-energy is $\epsilon^*$, given by
\begin{equation} 
\pi_1^2=(\epsilon^*)^2-\mu_1^2=(E_{\rm SR}^*)^2-m_1^2+O(\lambda^2).
\end{equation}
From this we immediately find
\begin{equation}
E_{\rm ISR}^*\approx E_{\rm SR}^*-\frac{\lambda}{2}((E_{\rm SR}^*)^2-m_1^2),
\end{equation}
which indicates a general lowering of threshold energies for $2\rightarrow N$ particle reactions.  The modified GZK threshold is simply the above result with $m_1 = m_p$.  Hence  Amelino-Camelia ISR also lowers the threshold so can not give an explanation of a possible raising of the GZK cutoff \cite{ACISR3}.  We note, however, that this result again depends on the assumption that the composite particle relations do not differ significantly from the SR relations.

We further illustrate the kinematics with the same processes studied before, V\v{C}R
and photon stability.  Both exact calculations are in Appendix B.  As in SR, there is no
V\v{C}R and the photon is stable in Amelino-Camelia ISR.

\section{On the Uniqueness of Process Thresholds}
\label{uniqueness}

The above results hold only if the map between the pseudo-variables and the physical variables is one-to-one.  If this property holds then there corresponds just one physical threshold for every threshold in special relativity. ISRs satisfy modified conservation laws in which the total energy-momentum
\begin{equation}
\begin{split} \label{econs}
E_{tot} &= F_\lambda(\epsilon_{tot},\pi_{tot}) \\
p_{tot} &= \pi_{tot} G_\lambda(\epsilon_{tot},\pi_{tot})
\end{split}
\end{equation}
are conserved. 

In this equation the total pseudo-energy-momenta
$(\epsilon_{tot},\pi_{tot})$ are functions of the physical
energy-momenta. For a single particle,
\begin{equation}
\begin{split} \label{pecons}
\epsilon= f_\lambda^{-1}(E,p) \\
\pi=p \, g_\lambda^{-1}(E,p)
\end{split}
\end{equation}
$f_\lambda$ and $g_\lambda$ may or may not be equivalent to
$F_\lambda$ and $G_\lambda$. For example, in Magueijo-Smolin ISR,
$F_\lambda=\epsilon w_\lambda(\epsilon)=f_\lambda$ and
$G_\lambda=w_\lambda(\epsilon)=g_\lambda$ with
$w_\lambda(\epsilon)=1/(1+\lambda \epsilon)$.  So in Magueijo-Smolin ISR the ``lower
case functions" are equivalent to ``upper case functions."

In the Amelino-Camelia ISR, however, the relevant equations are, for a single particle \cite{JV}
\begin{equation}
\begin{split}
E &= F_\lambda (\epsilon, \pi) = \frac{1}{\lambda} \ln \left[ \lambda \epsilon 
\cosh(\lambda m /2) +\cosh(\lambda m) \right] \\
p &= \pi G_\lambda (\epsilon, \pi) = \pi \, \cosh(\lambda m/2) e^{-\lambda E},
\end{split}
\end{equation}
and 
\begin{equation}
\begin{split}
\epsilon &= f_\lambda (E, p) = \frac{ e^{\lambda E} -\cosh(\lambda m)}{\lambda \cosh (\lambda m /2)} \\
\pi &= \pi g_\lambda (E, p) = \frac{p\,e^{\lambda E}}{\cosh(\lambda m/2)}
\end{split}
\end{equation}
which are simple inverses. 

In contrast to the single particle case for which $F_\lambda$ and $G_\lambda$ may be written as functions only of $\epsilon$ and $m$, in the multiple particle case the total energy and momentum are given by
\begin{equation}
\begin{split}
\label{ISR1FG}
F_\lambda(\epsilon,\pi)&=\frac{1}{\lambda} \ln \left[ 1+ \lambda \epsilon
\sqrt{1+\frac{\lambda^2}{4}(\epsilon^2-\pi^2)} +
\frac{\lambda^2}{2}(\epsilon^2-\pi^2) \right] \\
G_\lambda(\epsilon,\pi) &= \sqrt{ 1+ \frac{\lambda^2}{4} \left(
\epsilon^2-\pi^2 \right)}
\end{split}
\end{equation}
in which $\epsilon$ and $\pi$ are {\em sums} of the pseudo-energy-momentum variables for each particle. The functions are not identical; $F_\lambda \neq f_\lambda$ and
$G_\lambda \neq g_\lambda$.\footnote{These two expressions are equivalent for a single particle.  In the multiple particle case the problem arises because there is no longer a mass which relates the two expressions. Nevertheless, it
is easy to see that the expression $\epsilon^2-\pi^2$ is always
positive-semidefinite (zero in the case of a collection of photons).
For example, in the case of two particles from $|\epsilon_1| \geq |\pi_1|$
and $|\epsilon_2| \geq |\pi_2|$ it follows that the absolute value of the
sum $|\epsilon_1 + \epsilon_2|$ is also greater or equal than
$|\pi_1 + \pi_2|$ and so $\epsilon_{tot}^2 - \pi_{tot}^2 \geq 0$.
For more than two particles this can be generalized.}

Despite the apparent difference, the meaningful question is whether the mapping remains on-to-one.  Suppose $(E_o,p_o)$ is the total physical energy-momentum for the incoming particles obtained by summing the incoming particle pseudo-energy-momenta in equations (\ref{econs}). These modified energy conservation laws are equations for surfaces in energy-momentum space.  By the implicit function theorem, these surfaces determine solutions (generally, one-parameter families of solutions) only if the Jacobian of the functions is non-vanishing on their domain. More precisely, we require
\begin{equation}
\pi\left( \partial_\pi F_\lambda \partial_\epsilon G_\lambda -
\partial_\pi G_\lambda \partial_\epsilon F_\lambda \right) -
G_\lambda \partial_\epsilon F_\lambda \neq 0
\end{equation}
for $\epsilon \geq 0$ and $-\infty <\pi<\infty$.  The derivatives
are with respect to the pseudo-energy-momenta, e.g. $\partial_\pi
=\partial/\partial \pi$. For Magueijo-Smolin ISR this reduces to
\begin{equation}
-1/(1+\lambda \epsilon)^3 \neq 0.
\end{equation}
In the case of the Amelino-Camelia ISR, using equations (\ref{ISR1FG}) for the four dimensional case it is
\begin{equation}
-\frac{e^{-3\lambda E(\epsilon,\vec{\pi}) }}{1+\lambda^2( \epsilon^2-\vec{\pi}^{\:2} )/4}
\end{equation}
which is negative-definite, as well.\footnote{In the 1+1 case we find the Jacobian to be $$ e^{-\lambda E} \left[ \lambda^2(\epsilon^2-\pi^2)/4 \right] / 
\left[ 1+\lambda^2(\epsilon^2-\pi^2)/4 \right].$$}   Hence, both ISRs considered here have non-vanishing Jacobians and thus the mapping is bijective. The ISRs have no additional process thresholds.

\section{Discussion} \label{disc}

Using exact and first order calculations of process kinematics we have tested Amelino-Camelia ISR and Magueijo-Smolin ISR in their ``natural domain," momentum space.  Unlike previous kinematic calculations, these results made use of the Judes-Visser conservation laws \cite{JV}. The first order calculations in the CM frame show that Amelino-Camelia ISR lowers threshold energies, whereas the Magueijo-Smolin ISR may raise or lower threshold energies, for all allowed processes in special relativity.  The exact calculations exhibited in the Appendices show that there is no vacuum \v{C}erenkov radiation, forbidden in SR, and that photons are stable in these ISRs.  Finally, by studying the map to pseudo-energy-momentum variables we demonstrated that no processes beyond those in SR are allowed.

These results show that, when using the Judes-Visser modified conservation laws, the GZK threshold is {\em lowered} in these ISRs.  Although the ``GZK paradox" created by the apparent over-abundance of events above the GZK threshold is controversial \cite{waxman,demarco}, our analysis show that these ISRs do not provide a viable explanation of an apparent raising of the threshold. We note, however, that these results depend on both the form of the ISR energy-momentum conservation laws and the assumption on composite particles mentioned in Section \ref{ISR2}.

The kinematic results for the two example theories suggest two questions for any ISR: (i) Is the map between particle kinematic thresholds in the physical variables and the linear variables one-to-one? One source of trouble would be the existence of multiple threshold solutions which would require additional criteria to determine which solution is physical.  (ii) Are there processes normally forbidden in special relativity?  And at what energy and momentum do they occur?

In addition, in the ISR context we should expect covariance under
the modified transformations without requiring the energy-momenta
to take unphysical values. If agreement between observers requires
an unphysical boundary point of the physical state space then the
theory is not relativistic. 

These observations lead us to suggest sharpening the criteria of relativistic theories with an additional invariant scale. As in previous formulations of ISRs, (i) all modifications to special relativity must reduce to special relativity when the second invariant scale $\lambda$ ($E_p$) vanishes (diverges).  Physical solutions of the modified theories must reduce to the processes of special relativity in this limit.  Any theories which have multiple threshold solutions which satisfy this criteria are unphysical. (ii) Processes normally forbidden in special
relativity may only occur at the boundary (as determined by the additional scale) of the physical energy-momentum space. Therefore, ISRs can only shift processes (such as kinematic thresholds) or events but will not allow additional processes.

\begin{ack}
For discussions which illuminated key issues we thank Giovanni Amelino-Camelia, Tomasz Konopka, Jurek Kowalski-Glikman, Ralf Lehnert, Don Marolf, and Lee Smolin. We also thank Julien LeBrun of Pierre et Marie Curie University, Paris  for performing linear calculations. D.H. was supported, in part, by the Ralph E. Hansmann Science Students Support and the Casstevens Scholarship Funds of Hamilton College. F.H. would like to acknowledge Hamilton College for hospitality and the Czech ministry of education for  support (contract no. 143100006). S.M. acknowledges the Perimeter Institute for hospitality and the Research Corporation for support. 
\end{ack}

\section{Appendix A}

\noindent {\bf Boost for GZK threshold} To find the boost from the CM frame to the cosmological frame one can use the CM condition
\begin{equation}
\frac{P_p}{1-\lambda E_p} = - \frac{P_\gamma}{1-\lambda E_\gamma}
\end{equation}
to find $E_\gamma$, the energy of the photon in the CM frame.  Boosting this energy to give $\epsilon$, the energy of the far infrared photon in the cosmological frame gives $\gamma$
\begin{equation}
\gamma = \frac{E_\gamma^2+\epsilon^2-2E_\gamma^2 \epsilon \lambda- 2\lambda E_\gamma \epsilon^2 + 2 \lambda^2 \epsilon^2 E_\gamma^2}{2 E_\gamma \epsilon (1- \lambda E_\gamma)(1-\lambda \epsilon)}.
\end{equation}
With the modified dispersion relation, equation (\ref{SRGZKCMthresh}), and the equation for $E_\gamma$ it is possible to use the above $\gamma$ to boost the threshold back into the cosmological frame.  The result, to leading order in $\lambda$  (with $m\equiv m_p$) is 
\begin{equation}
\begin{split}
E_{\rm ISR}^*&\approx \frac{4 {\epsilon}^2 m^2 + {m_\pi}^2 
      {\left( 2 m + m_\pi \right) }^2}{4  \epsilon m_\pi 
     \left( 2 m + m_\pi \right) }  - \lambda \left[ {m_\pi}^4 
        \left( m + m_\pi \right)   {\left( 2 m + m_\pi \right) }^4 \right.   \\ 
     &+ \left. 16 {\epsilon}^3 m^2 \left\{ \epsilon \left( m^3 - m^2 m_\pi -
          3 m {m_\pi}^2 - {m_\pi}^3\right) - m_\pi  \left( 6 m^3 + 8 m^2 m_\pi + 2 m {m_\pi}^2 - {m_\pi}^3
          \right)  \right\} \right. \\
          &- \left. 4 \epsilon {m_\pi}^3 {\left( 2 m + m_\pi \right) }^2 \left( -2 m^3 - 2 m^2 m_\pi + {m_\pi}^3 \right)  + 4 {\epsilon}^2 {m_\pi}^2  {\left( 2 m + m_\pi \right) }^2 \right. \\ 
       &\times \left. \left( 2 m^3 + 4 m^2 m_\pi + 3 m {m_\pi}^2 + {m_\pi}^3
          \right)  \right] / 16 {\epsilon}^2  {m_\pi}^2 
     \left( m + m_\pi \right)   \left( 2 m + m_\pi \right)^2
     \end{split}
\end{equation}
Expanding this in leading terms assuming $m_\pi/m \ll1$ and $\epsilon/m_\pi \ll 1$ one finds that
\begin{equation}
E_{\rm ISR}^*\approx \frac{m m_\pi}{2 \epsilon} - \lambda \left( \frac{m m_\pi}{2 \epsilon} \right)^2 = E_{\rm ISR}^* - \lambda (E^*_{\rm SR})^2 
\end{equation}
so, not surprisingly, the boost modifications swamp the mass modifications.

\noindent {\bf V\v{C}R} Vacuum \v{C}erenkov radiation (V\v{C}R) may occur in theories with
modified dispersion relations, and indeed this process places
strong limits on the extent of the modification \cite{KM}. Since
ISRs apparently do not require a preferred frame we can make
use of the usual process kinematics techniques of SR. In the rest
frame of the incoming charged particle let the energy-momentum be
$(E_o,p_o)=(m_a,0)$. We denote the products' energy momenta as
$(E_a, p_a)$ and $(E_\gamma, p_\gamma)$. The modified conservation
of momentum immediately gives $\pi_a = - \pi_\gamma$. The modified
conservation of energy is then
\begin{equation}
\begin{split}
E_o=E_{tot}=\frac{\epsilon_a
+\epsilon_\gamma}{1+\lambda(\epsilon_a +\epsilon_\gamma)} \\
= \frac{\epsilon_a - \pi_a}{1+\lambda(\epsilon_a - \pi_a)}
\end{split}\end{equation}
With the dispersion relation $(\epsilon_a
-\pi_a)(\epsilon_a+\pi_a)=\mu_a^2$ one can re-express energy
conservation as a simple polynomial in $\epsilon_a$ which has but
one solution $(\epsilon_a,\pi_a)=(\mu_a,0)$.  Therefore since the
photon's physical momentum vanishes, V\v{C}R does not occur.

\noindent {\bf Photon Stability} In the case of photon stability we use a different method that
does not require a choice of reference frame. We denote the photon
energy-momentum by $(E_\gamma,p_\gamma)$ and the electron-positron
pair energy-momenta by $(E_\pm,p_\pm)$. In Magueijo-Smolin ISR, the
pseudo-momentum is conserved so we have
$\epsilon_{tot}=\epsilon_\gamma = \pi_\gamma$ with the last
equality being true for massless particles.  The relation gives
the simple result,
\begin{equation}
\frac{E_+}{1-\lambda E_+} - \frac{p_+}{1-\lambda E_+} =
-\frac{E_-}{1-\lambda E_-} + \frac{p_-}{1-\lambda E_-}.
\end{equation}
With the energy and momentum of the outgoing particles separated
we simply need to understand the behavior of one function.  Using
the dispersion relations of equation (\ref{ISR2disp}) we simply
have
\begin{equation} \label{oddness}
f(E_+)=-f(E_-) \end{equation} with \begin{equation} f(E)=
\frac{E-\sqrt{E^2-m^2 \left( \frac{1-\lambda E}{1-\lambda m}
\right)^2}}{1-\lambda E}.  \end{equation} The condition of equation
(\ref{oddness}) is only satisfied at a root of $f(E)=0$.  However, this
only occurs when $E=E_p$.  Since this point is excluded, the photon is
stable.

\section{Appendix B}

\noindent {\bf V\v{C}R} The vacuum \v{C}erenkov calculation proceeds as in Magueijo-Smolin ISR when one takes the rest frame of the incoming charged particle.  In Amelino-Camelia ISR, however, the modified energy
conservation becomes,
\begin{equation}
\label{ISR1vcrecons}
 m_a = \frac{1}{\lambda} \ln \left[ 1 +
\lambda \epsilon_{tot} \sqrt{ 1+ \frac{\lambda^2
\epsilon_{tot}^2}{4}} + \frac{\lambda^2}{2} \epsilon_{tot}^2
\right]
\end{equation}
with
\begin{equation}
\epsilon_{tot}= \frac{e^{\lambda E_a} -\cosh(\lambda m_a)}{\lambda
\cosh(\lambda m_a/2)}+\frac{e^{\lambda E_\gamma}-1}{\lambda}.
\end{equation}
The expression of equation (\ref{ISR1vcrecons}) simply gives,
after a bit of algebra,
\begin{equation}
\epsilon_{tot} = \frac{2 \sinh(\lambda m_a /2)}{\lambda} \equiv
\mu_a.
\end{equation}
Since the pseudo-energy is equivalent to the pseudo-mass it is not
surprising that we find, from the definition of $\epsilon_{tot}$,
that $E_\gamma =0$ and $(E_a,p_a)=(m_a,0)$.  As in SR, there is no
V\v{C}R in Amelino-Camelia ISR.

\noindent {\bf Photon Stability} In the Amelino-Camelia ISR, conservation of energy $E_\gamma = E_{tot}$ gives
\begin{equation}
\epsilon_\gamma = \epsilon_{tot} \sqrt{ 1+
\frac{\lambda^2(\epsilon_{tot}^2-\pi_{tot}^2)}{4}} +
\frac{\lambda(\epsilon_{tot}^2 - \pi_{tot}^2)}{2}.
\end{equation}
But photons have the property that
$\epsilon_\gamma^2=\pi_\gamma^2$.  So we can use momentum
conservation $p_\gamma = p_{tot}$ to simplify this.  In fact,
\begin{equation}
\epsilon_\gamma^2 = \pi_{tot}^2 \left( 1 + \frac{\lambda^2}{4}
(\epsilon_{tot}^2 - \pi_{tot}^2) \right).
\end{equation}
Equating the two expressions for $\epsilon_\gamma^2$ we have the
result
\begin{eqnarray} \label{photcons}
0&=& (\epsilon_{tot}^2 - \pi_{tot}^2)\left[ 1 + \frac{\lambda^2}{2}
(\epsilon_{tot}^2 - \pi_{tot}^2) + \lambda \epsilon_{tot} \sqrt{ 1
+ \frac{\lambda^2}{4}(\epsilon_{tot}^2 - \pi_{tot}^2)} \right]\nonumber\\
&=&(\epsilon_{tot}^2-\pi_{tot}^2)\,e^{\lambda E_{tot}}.
\end{eqnarray}
The first solution to equation (\ref{photcons}), when the first
factor vanishes, gives $E=-m$. This is the result that one would
obtain in SR by an analogous calculation. Since $E>0$, the
`solution' is unphysical. For the same reason the second factor cannot vanish.
Hence, there are no massive-particle solutions, so the photon is stable in the Amelino-Camelia framework as well.

\end{document}